\documentclass[twocolumn,showpacs,prl]{revtex4}

\usepackage{amsmath,amssymb}
\usepackage{graphicx}
\usepackage{verbatim}

\draft
\begin{document}

\title{Effect of Induced Spin-Orbit Coupling for Atoms via Laser Fields}

\author{Xiong-Jun Liu, Mario F. Borunda, Xin Liu and Jairo Sinova}
\affiliation{Department of Physics, Texas A\&M University, College
Station, Texas 77843-4242, USA}

\begin{abstract}
We propose an experimental scheme to observe spin-orbit coupling
effects of a two-dimensional (2D) Fermi atomic gas cloud by
coupling its internal electronic states (pseudospins) to radiation
in a Lambda configuration. The induced spin-orbit (SO) coupling
can be of the Dresselhaus and Rashba type with and without a
Zeeman term. We show that the optically induced SO coupling can
lead to a spin-dependent effective mass under appropriate
condition, with one of them able to be tuned between positive and
negative effective masses. As a direct observable we show that in
the expansion dynamics of the atomic cloud the initial atomic
cloud splits into two clouds  for the positive effective mass case
regime, and into four clouds for the negative effective mass
regime.
\end{abstract}
\date{\today }
\maketitle

\indent Spin-orbit (SO) coupling effect in semiconductors has
emerged in the solid-state community as a very active field of
research, fuelled in part by the field of spintronics
\cite{spintronics}, e.g. the engineering of devices where the spin
degree of freedom of the electron is exploited for improved
functionality. This has lead to new developments in the anomalous
Hall effect (AHE) \cite{AHE} and the spin Hall effect (SHE)
\cite{SHE1,SHE2}. In correspondence to the spin of an electron,
the internal degree of freedom of an atom (pseudospin) is
represented by the superposition of its electronic states
(hyperfine levels). SO coupling can be equivalently depicted as
the interaction between an effective non-Abelian gauge potential
and a particle with (pseudo)spin. In quantum systems, the idea
generating a gauge field adiabatically was proposed by Wilczek and
Zee more than twenty years ago \cite{wilczek}. Recently, such an
idea was applied to atomic systems, where the motion of atoms in a
position dependent laser configuration gives rise to an effective
non-Abelian gauge potential
\cite{atomgauge1,atomgauge2,atomgauge3,atomgauge4,atomgauge5},
which can lead to an effective SO interaction in an ultracold
atomic gas \cite{spin-orbit,Clark,niu1}.

Realization of SO interaction in atomic gases opens new
possibility of studying spintronic effects, e.g. spin relaxation
\cite{spin-orbit}, $Zitterbewegung$ \cite{Clark} and SHE, in
atomic systems which provide an extremely clean environment,
allowing in a controllable fashion unique access to the study of
complex physics. However, experimental detection of such SO
effects in atoms requires to measure the pseudospins (not just
hyperfine levels) that are usually not directly observable for
atomic systems. In this letter, we propose an experimental scheme
to study SO coupling effects, based on a trapped two-dimensional
(2D) Fermi atomic gas with a simple internal three-level
$\Lambda$-type setup. We demonstrate that an effective SO
interaction, e.g. Rashba and linear Dresselhaus terms, can be
obtained by coupling atoms with a three-level configuration to
spatially varying laser fields. The optically induced SO coupling
can lead to a spin-dependent effective masses under proper
condition.  A direct observable of this effects is in the
expansion dynamics for each of the effective mass cases after the
external trap is switched off and we predict that the initial
atomic cloud splits into two or four clouds.

\begin{figure}[ht]
\includegraphics[width=0.9\columnwidth]{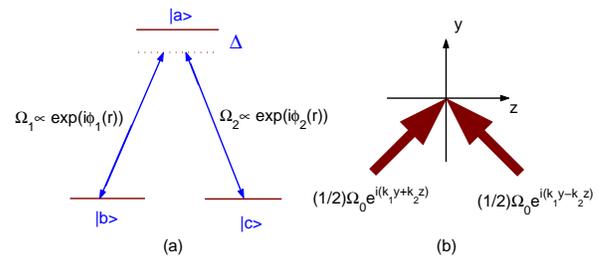}
\caption{(Color online) (a) Three-level $\Lambda$-type system
coupled to position-dependent laser fields with large detuning;
(b) Laser configuration for $\Omega_1$.} \label{fig1}
\end{figure}
We consider a cloud of quasi 2D ($y$-$z$ plane) Fermi atomic gas
with internal three-level $\Lambda$-type configuration (see Fig. 1
(a)) coupled to radiation. The transition
$|b\rangle\rightarrow|a\rangle$ is coupled by the laser field with
Rabi-frequency $\Omega_1=\Omega_{10}\exp[i\phi_1(\bold r)]$ and
the transition $|c\rangle\rightarrow|a\rangle$ is coupled by
another laser field $\Omega_2=\Omega_{20}\exp[i\phi_2(\bold r)]$,
where $\phi_{1,2}(\bold r)$ are position-dependent phases. The
Hamiltonian of a single particle reads: $H=H_0+V_{trap}(\bold
r)+H_I$, where $V_{trap}(\bold r)$ is the 2D harmonic trap, and
the interacting Hamiltonian is given by
\begin{eqnarray}\label{eqn:H5}
H_I=\hbar\Delta|a\rangle\langle a|-(\hbar\Omega_1|a\rangle\langle
b|+\hbar\Omega_2|a\rangle\langle c|+h.c.).
\end{eqnarray}
Diagonalizing this Hamiltonian yields three eigenstates:
$|\chi_D\rangle=\sin\theta|b\rangle-\cos\theta
e^{i\phi}|c\rangle$,
$|\chi_{B_1}\rangle=\cos\alpha\cos\theta|b\rangle+\cos\alpha\sin\theta
e^{i\phi}|c\rangle+\sin\alpha e^{i\phi_1}|a\rangle$, and
$|\chi_{B_2}\rangle=\sin\alpha\cos\theta|b\rangle+\sin\alpha\sin\theta
e^{i\phi}|c\rangle-\cos\alpha e^{i\phi_1}|a\rangle$. Here
$\phi=\phi_1-\phi_2$, the mixing angles
$\tan\theta=\Omega_{20}/\Omega_{10}$ and
$\tan\alpha=\sqrt{\Omega_{10}^2+\Omega_{20}^2}/\Delta\equiv\Omega_0/\Delta$.
The corresponding eigenvalues are $E_{D}=0$ and
$E_{B_{1,2}}=\hbar(\Delta\mp\sqrt{\Delta^2+4\Omega_{0}^2})/2$.
Since spatially-varying lasers are employed, diagonalization of
the interacting Hamiltonian $H_I$ leads to a SU(3) gauge potential
\cite{atomgauge1,atomgauge2,atomgauge3,atomgauge4}. We consider
the large detuning case, $\Delta^2\gg\Omega_0^2$, where
$|E_{D}-E_{B_1}|\ll\Omega_0$ and
$|\chi_{D}\rangle,|\chi_{B_1}\rangle$ spans a near-degenerate
subspace, with their eigenvalues far separated from that of
$E_{B_2}$. We can then apply the adiabatic condition by neglecting
the state $|\chi_{B_2}\rangle$, which leads to a U(2) non-Abelian
adiabatic gauge potential based on the near-degenerate subspace
spanned by $|\chi_{D,B_1}\rangle$ \cite{note}. This situation is
different from the cases in Refs. \onlinecite{atomgauge4} and
\onlinecite{spin-orbit} where the adiabatic condition is also
assumed between the states $|\chi_D\rangle$ and
$|\chi_{B_1}\rangle$, and thus the spin-dependent gauge potential
is still Abelian in their case \cite{atomgauge4}. The adiabatic
non-Ablelian gauge potential for the present near-degenerate
subspace is obtained via
\begin{eqnarray}\label{eqn:gauge10}
\bold A(\bold r)=i\frac{\hbar c}{e}\langle \chi_D|\otimes\langle
\chi_{B_1}|\nabla|\chi_{B_1}\rangle\otimes|\chi_D\rangle.
\end{eqnarray}
For our purpose, we shall set the parameters $\phi_1=\phi_2=k_1y$
and $\theta=k_2z$, which means $\Omega_{1,2}$ are standing waves
in the $z$ direction but plane waves in the $y$ direction. Such
configuration can be achieved by applying two laser fields for
each atomic transition. For $\Omega_1$, for instance, we can set
two laser fields with the same strength, where one travels with
the wave vector $k_1\hat e_y+k_2\hat e_z$ and the other travels
with $k_1\hat e_y-k_2\hat e_z$ (see Fig. 1 (b)), similarly can be
done for $\Omega_2$. The Rabi-frequencies are then given by
$\Omega_1(\bold r)=\Omega_{0}\cos(k_2z)e^{ik_1y}$ and
$\Omega_2(\bold r) =\Omega_{0}\sin(k_2z)e^{ik_1y}$. For
simplicity, in what follows, we use the spin language and denote
by $|\uparrow\rangle=|\chi_D\rangle$,
$|\downarrow\rangle=|\chi_{B_1}\rangle$. Under this configuration
the gauge field (\ref{eqn:gauge10}) can be recast into $\bold
A=m\lambda_1\sigma_z\hat e_y-m\lambda_2\sigma_y\hat
e_z-m\lambda_1I\hat e_y$ with the coefficients $\lambda_1=\hbar
k_1\Omega_0^2/(2m\Delta_0^2)$ and $\lambda_2=\hbar k_2/m$ (note
$\bold A$ does not depend on mass $m$, and its current form is for
the definition of $\lambda_{1,2}$). In addition, the scalar
potentials are given by
$\varphi_{\uparrow\uparrow}=\frac{\hbar^2}{2m}\frac{\Omega_0^2}{\Delta^2}k_2^2$,
$\varphi_{\downarrow\downarrow}=\frac{\hbar^2}{2m}\frac{\Omega_0^2}{\Delta^2}k_1^2$,
and $\varphi_{\downarrow\uparrow}=0$.

To this step we can obtain the effective Hamiltonian for the
near-degenerate subspace with SO coupling in the form (neglecting
constant terms):
\begin{eqnarray}\label{eqn:effectiveHamiltonian1}
H=H_0+H_{so}+H_{z}+V_{trap},
\end{eqnarray}
where $H_0=\frac{P_y^2}{2m}+\frac{P_z^2}{2m}$,
$H_{so}=-\lambda_1\sigma_zP_y+\lambda_2\sigma_yP_z$,
$H_{z}=M_0\sigma_z$ with
$M_0=\frac{\hbar^2}{4m}\frac{\Omega_0^2}{\Delta^2}k_2+\hbar\frac{\Omega_0^2}{2\Delta}$,
and the 2D harmonic trap $V_{trap}=\frac{1}{2}m\omega^2(y^2+z^2)$.
A Hamiltonian of this form is predicted to give SHE \cite{SHE1}
for $M_0=0$ and planar Hall effect \cite{PHE} for $M_0\neq0$ in
solid state systems.

The term $H_{so}+H_{z}$ can be readily diagonalized in the
momentum space, $(H_{so}+H_{z})^{diag}
=\hbar\sqrt{(\frac{M_0}{\hbar}-\lambda_1k_y)^2+\lambda_2^2k_z^2}\sigma_z$,
where $k_{y,z}$ are wave vectors of atoms in the $y$ and $z$
directions. The associated eigenstates are given by
$|+\rangle=[\cos\vartheta/2, i\sin\vartheta/2]^{T}$,
$|-\rangle=[i\sin\vartheta/2, \cos\vartheta/2]^{T}$ with
$\tan\vartheta=\lambda_2\hbar k_z/(M_0-\lambda_1\hbar k_y)$.
However, even in this situation, the full Hamiltonian $H$ is not
diagonalizable. Practically, we can consider the case where
$M_0^2\gg\lambda_2^2\hbar^2(k_F^z)^2\gg\lambda_1^2\hbar^2(k_F^y)^2$
(the validity in realizable experimental set-ups will be discussed
below), $k_F^{y,z}$ are the $y/z$-components of the Fermi momenta,
and expand the term $(H_{so}+H_{z})^{diag}$ to the $k^2$ order:
$(H_{so}+H_{z})^{diag}
\approx(M_0+\frac{\lambda_2^2P_z^2}{2M_0}-\lambda_1P_y)\sigma_z$.
Substituting this result into eq.
(\ref{eqn:effectiveHamiltonian1}) yields
\begin{eqnarray}\label{eqn:effectiveHamiltonian3}
H^{(1)}&=&\frac{1}{2m}(P_y-\lambda_1m\sigma_z)^2+\frac{1}{2\tilde{m}}P_z^2+M_0\sigma_z\nonumber\\
&& +\frac{1}{2}m\omega^2\bold r^2-\frac{1}{2}m\lambda_1^2,
\end{eqnarray}
where
$\frac{1}{\tilde{m}}=\frac{1}{m}+\frac{\lambda_2^2}{M_0}\sigma_z$.
Eq. (\ref{eqn:effectiveHamiltonian3}) shows that the SO coupling
leads to a spin-dependent effective mass of atoms moving in the
$z$ direction, with $\tilde{m}_\pm=(1\pm\delta m/m)^{-1}m$ and
$\delta m=\hbar^2k_2^2/M_0$. The eigenfunction of Eq.
(\ref{eqn:effectiveHamiltonian3}) reads
\begin{eqnarray}\label{eqn:eigenfunction1}
\Psi_{n_y,n_z}^\eta(y,z)=\psi_{n_y}(\omega_y,y)\tilde{\psi}_{n_z}^{\eta}
(\omega_z^\eta,z)e^{i\eta\lambda_1my/\hbar}|\eta\rangle,
\end{eqnarray}
with $|\eta\rangle=|\pm\rangle$. Here
$\psi_{n_y},\tilde{\psi}_{n_z}^{\eta}$ are harmonic oscillator
wave functions. The eigenvalues are given by
$E_{n_y,n_z}^\eta=(n_y+\frac{1}{2})\hbar\omega_y+(n_z+\frac{1}{2})\hbar\omega_z^{\eta}+\eta
M_0-\frac{1}{2}m\lambda_1^2$, where $\omega_y=\omega$,
$\omega_z^{\eta}=\omega\sqrt{1+\eta\delta m/m}$ are spin-dependent
effective trap frequencies due to SO coupling, and $n_y, n_z$ are
integers.

The results (\ref{eqn:effectiveHamiltonian3}) and
(\ref{eqn:eigenfunction1}) are valid for the situation $\delta
m<m$, where the effective mass of particles in state $|-\rangle$
is positive, i.e. $\tilde{m}_->0$. On the other hand, the negative
effective mass regime for the $|-\rangle$ state can be reached
when $\delta m>m$. In this case one can verify the dispersion
relation $\varepsilon_{k}^-=H_0+H^{(-)}_{so}+H^{(-)}_{z}$ (for
atoms in $|-\rangle$) represents a double-well potential in the
$k_z$ axis, therefore a higher-order expansion with momentum $k_z$
is required to derive the effective Hamiltonian. Similarly, we
consider that $\lambda_2^4\hbar^4(k_F^z)^4\ll M_0^4$ and can then
expand $\varepsilon_{k}^-$ up to the $k_z^4$ order, which gives a
double well $\phi^4$-type potential form in $k_z$-momentum space
($\tilde{m}_-<0$), i.e.
$\varepsilon_{k_z}^-=\frac{\lambda_2^4}{8M_0^3}P_z^4+\frac{1}{2\tilde{m}_-}P_z^2$
\cite{doublewell}. Higher-order terms in the expansion can
equivalently lead to small corrections to the coefficients of
$k_z^2$ and $k_z^4$ terms. Finally we get effective Hamiltonian
for atoms at state $|-\rangle$:
\begin{eqnarray}\label{eqn:effectiveHamiltonian5}
H^{(2)}_-&=&\frac{1}{2m}(P_y+\lambda_1m)^2+\frac{\lambda_2^4}{8M_0^3}(1+\gamma_1)P_z^4
\nonumber\\
&+&\frac{1}{2\tilde{m}_-}(1+\gamma_2)P_z^2-M_0
+\frac{1}{2}m\omega^2\bold r^2-\frac{1}{2}m\lambda_1^2.
\end{eqnarray}
Here the coefficient correction $\gamma_1=\frac{4}{\frac{\delta
m}{m}(1+\frac{\delta m}{m})^2}-1, \gamma_2=\frac{2}{1+\frac{\delta
m}{m}}-1$ are small under the condition
$\lambda_2^4\hbar^4(k_F^z)^4\ll M_0^4$, say, $\gamma_{1,2}^2\ll1$
in the present case.

\begin{figure}[ht]
\includegraphics[width=0.65\columnwidth]{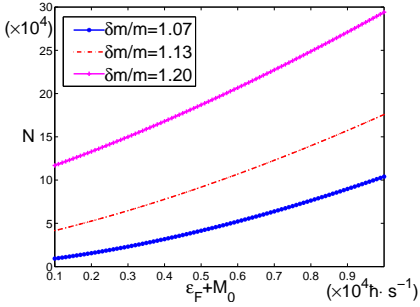}
\caption{(Color online) Number of atoms versus Fermi energy at
$\delta m/m=1.07, 1.13, 1.20$, corresponding to the effective mass
$\tilde{m}_-=-15m, -7.7m, -5m$.} \label{fig2}
\end{figure}

Unlike the positive mass regime, calculation of the Fermi energy
in the negative mass case cannot be done analytically.
Nevertheless, we are interested in systems with a large number of
atoms, where the Thomas-Fermi (TF) approximation is suitable
\cite{fermivelocity,40K,Granade}. Then the Fermi energy can be
calculated by solving the equation:
\begin{eqnarray}\label{eqn:fermi5}
N=\sum_{\eta=+,-}\int_{k<k^{\eta}_F,r<R_F^\eta} d^2\bold rd^2\bold
k\rho^{(2)}_\eta(\bold r,\bold k,T=0),
\end{eqnarray}
where the atomic distribution function in phase space is given by
$\rho^{(2)}_\pm=(2\pi)^{-2}(e^{\beta(H^{(2)}_\pm(\bold r,\bold
k)-\mu_F)}+1)^{-1}$ with $\beta=1/k_BT$, and the initial size of
the atomic cloud is given by $R_F^{\pm}=(2\varepsilon_F\mp
2M_0-2\min\{\varepsilon_{k_z}^\pm\})^{1/2}/(m^{1/2}\omega)$. From
Eq. (\ref{eqn:fermi5}) the relation of Fermi energy to the number
of atoms $N$, trap frequency and SO coupling strength can be
obtained numerically as shown in Fig.\ref{fig2}.

There are specific observables of this spin-dependent effective
mass induced by the spin-orbit coupling. Firstly, the anisotropy
of the effective mass can lead to the anisotropic momentum
distribution. Considering the Thomas-Fermi approximation, at zero
temperature and for the positive mass case, the momentum
distribution is given by $n_\pm(\bold k)=\frac{1}{2\pi
m\omega^2}\bigr[\varepsilon_F\mp
M_0-\frac{\hbar^2}{2m}(k_y^2+\frac{m}{\tilde{m}_\pm}k_z^2)\bigr]$.
This contrasts with the result for the usual isotropic mass, where
the atomic distribution in the momentum space is always isotropic
irrespective of the shape of the trap \cite{fermivelocity}. For
the negative effective mass case, $n_+(\bold k)$ has the same
form, while $n_-(\bold k)=\frac{1}{2\pi
m\omega^2}\bigr[\varepsilon_F
+M_0-\frac{\hbar^2}{2m}(k_y^2+\frac{m}{\tilde{m}_-}(1+\gamma_2)k_z^2
+\frac{m\lambda_2^4\hbar^2}{4M_0^3}(1+\gamma_1)k_z^4)\bigr]$.
Also, a fully polarized Fermi gas is obtained when
$\varepsilon_F<M_0$, and $n_+(\bold k)=0$. The anisotropy of the
momentum distribution can lead to anisotropy of the Fermi velocity
in the $y$-$z$ plane. For instance, when $\delta m/m=3/4$ and
$\varepsilon_F<M_0$, we have $v^z_F\approx v^y_F/2$. The
anisotropy of Fermi velocity can be directly detected by
time-of-flight absorption \cite{40K,Granade}.

However, the dramatic signature of SO effects in the present Fermi
atomic gas resides in the expansion dynamics of the atomic cloud
after the 2D external trap and the laser fields are switched off.
The evolution of the atomic distribution in phase space
$\rho(\bold r,\bold k,T,t)$ can be calculated by the Boltzmann
transport equation assuming that at $t=0$, $V_{trap}\rightarrow0$
and $M_0\rightarrow0$. For the present cold dilute non-interacting
Fermi gas, the evolution of $\rho_{b,c}(\bold r,\bold k,T,t)$ is
followed by the ballistic law $\rho_{b,c}(\bold r,\bold
k,T,t>0)=\rho_{b,c}(\bold r-\hbar\bold kt/m,\bold k,T,0)$ (here
$|\bold k, \nu\rangle, (\nu=b,c)$ are the eigenstates after the
trap and laser fields are turned off). The temporal atomic spatial
density can be calculated by: $n_{b,c}(\bold r,T,t)=\int d^2\bold
k\rho_{b,c}(\bold r-\hbar\bold kt/m,\bold k,T,0)$. The eigenstates
$|\bold k, \alpha\rangle, (\alpha=b,c)$ are related to the initial
pseudospin basis $|\pm\rangle$ by $|\bold
k,b\rangle\approx\frac{1}{2}|\bold k+k_2\hat
e_z,+\rangle+\frac{1}{2}|\bold k-k_2\hat
e_z,+\rangle+\frac{1}{2i}|\bold k+k_2\hat
e_z,-\rangle-\frac{1}{2i}|\bold k-k_2\hat e_z,-\rangle$, and
$|\bold k,c\rangle\approx\frac{1}{2}|\bold k+k_2\hat
e_z,-\rangle+\frac{1}{2}|\bold k-k_2\hat
e_z,-\rangle-\frac{1}{2i}|\bold k+k_2\hat
e_z,+\rangle+\frac{1}{2i}|\bold k-k_2\hat e_z,+\rangle$. We find
that the atomic density is:
\begin{eqnarray}\label{eqn:expansion4}
n_{b}(\bold r,T,t)&=&\sum_{\eta=\pm}\int\frac{d^2\bold k}{4}
\biggr\{\rho_\eta^{(j)}[\bold r-\frac{\hbar t}{m}(\bold k+\eta k_2\hat e_z),\bold k,T,0]\nonumber\\
&+&\rho_\eta^{(j)}[\bold r-\frac{\hbar t}{m}(\bold k-\eta k_2\hat
e_z),\bold k,T,0]\biggr\},\end{eqnarray} where
$\rho_\pm^{(j)}(\bold r,\bold k,T,0)$ denote the distribution
functions for atoms in the state $|\pm\rangle$ for the positive
mass ($j=1$) and negative mass ($j=2$) cases. The function
$n_c(\bold r,T,t)$ can be calculated in the same way. Practically,
we can assume that before the expansion begins
$\varepsilon_F<M_0$, and then $\rho_+^{(j)}(r,\bold k,T,0)=0$. For
the positive mass case, the evolution of the atomic density can be
calculated exactly:
\begin{eqnarray}\label{eqn:momentumfunction3}
n_b(\bold
r,T,t)&=&\frac{\sqrt{m\tilde{m}_-}}{8(1+\omega^2t^2)\beta\pi\hbar^2}
\bigr(\frac{1+\omega^2t^2}{1+\frac{\tilde{m}_-}{m}\omega^2t^2}\bigr)^{1/2}\\
&&\times\bigr(\ln\frac{1+e^{-\beta \tilde{E}_+(\bold
r,t)}}{e^{-\beta \tilde{E}_+(\bold r,t)}}+\ln\frac{1+e^{-\beta
\tilde{E}_-(\bold r,t)}}{e^{-\beta \tilde{E}_-(\bold
r,t)}}\bigr),\nonumber
\end{eqnarray}
where $\tilde{E}_\pm(\bold r,t)=\mu_F+M_0-\frac{m\omega^2y^2}
{2(1+\omega^2t^2)}-\frac{m\omega^2(z\pm\hbar
k_2t/m)^2}{2(1+\frac{\tilde{m}_-}{m}\omega^2t^2)}$. One can verify
that at $t=0$ the maximum point of $n_b$ is obtained at $\bold
r=0$, while after a sufficiently long time there are two maximum
points at $y=0,z=\pm\hbar k_2t/m$. As a result, Eq.
(\ref{eqn:momentumfunction3}) represents an initial atomic cloud
that splits into two clouds each moving in opposite direction with
group velocities $v_g=\pm\hbar k_2\hat e_z/m$. Using typical
parameters: $M_0=10^{6}\hbar\cdot s^{-1},
m\approx0.963\times10^{-26}kg$ ($^6$Li atoms) and
$k_2=0.87\times10^{7}$m$^{-2}$, one finds
$\tilde{m}_-\approx4m$, and $v_g\approx\pm9.0cm/s$.

The evolution of the atomic density in the negative effective mass
regime is more complicated. Again, assuming $\varepsilon_F<M_0$,
we find from Eq. (\ref{eqn:expansion4}) the temporal atomic
density as (denoting by $\tilde{z}_\eta=z\pm\hbar k_2t/m$):
\begin{eqnarray}\label{eqn:expansion18}
&n_b&(\bold r,T,t)=\sum_{\eta=+,-}\int\frac{d^2\bold k}{16\pi^2}
\biggr\{1+\exp\bigr\{\beta[\frac{1+\omega^2t^2}{2m}\hbar^2
k^2_y\nonumber\\
&&+\frac{\Gamma}{2m\hbar^2}\bigr(\hbar
k_z+\frac{\tilde{m}_-\omega^2t\tilde{z}_\eta}{1+\frac{\tilde{m}_-}{m}\omega^2t^2}\bigr)^4
+\frac{m+\tilde{m}_-\omega^2t^2}{2m\tilde{m}_-}\hbar^2
k^2_z\nonumber\\
&&+\frac{m\omega^2y^2}{2+2\omega^2t^2}+\frac{m^2\omega^2\tilde{z}_\eta^2}{2m+2\tilde{m}_-\omega^2t^2}
-M_0-\mu_F]\bigr\}\biggr\}^{-1},
\end{eqnarray}
where $\Gamma=m\lambda_2^4\hbar^2(1+\gamma_1)/(4M_0^3)$. To make a
qualitative analysis, we first calculate the time independent
momentum distribution: $n_b(\bold k,T,t)=\frac{1}{8\pi
m\omega^2}(\ln\frac{1+e^{-\beta \tilde{\cal E}_+(\bold
k)}}{e^{-\beta \tilde{\cal E}_+(\bold k)}}+\ln\frac{1+e^{-\beta
\tilde{\cal E}_-(\bold k)}}{e^{-\beta \tilde{\cal E}_-(\bold
k)}})$. Here $\tilde{\cal E}_\pm(\bold
k)=\mu_F+M_0+\frac{\hbar^2\Gamma\Upsilon^2}{2m}-
\frac{\hbar^2}{2m}\bigr[k_y^2+\Gamma\bigr((k_z\pm
k_2)^2-\Upsilon\bigr)^2\bigr]$ with
$\Upsilon=-m(1+\gamma_2)/(2\tilde{m}_-\Gamma)$. Note that if
$k_2\neq\sqrt{\Upsilon}$, $n_b(\bold k,T,t)$ has four distinct
maximums at: $k_z=-k_2-\sqrt{\Upsilon}, \ -k_2+\sqrt{\Upsilon}, \
k_2-\sqrt{\Upsilon}, \ k_2+\sqrt{\Upsilon}$, which indicates that
the initial atomic cloud is composed of four overlapping clouds
that will travel in the $z$-axis at different speeds. Using
parameters similar to the previous system:
$m\approx0.963\times10^{-26}kg$ ($^6$Li atoms),
$k_2=1.0\times10^7$m$^{-1}$, and $M_0=0.90\times10^{6}\hbar\cdot
s^{-1}$, we find that $\Gamma\approx4.43\times10^{-15}$m$^2$ and
$\Upsilon=8.9\times10^{12}$m$^{-2}$. Thus the four maximums
correspond to $v_g=\pm13.4cm/s, \pm7.4cm/s$.

\begin{figure}[ht]
\includegraphics[width=0.6
\columnwidth]{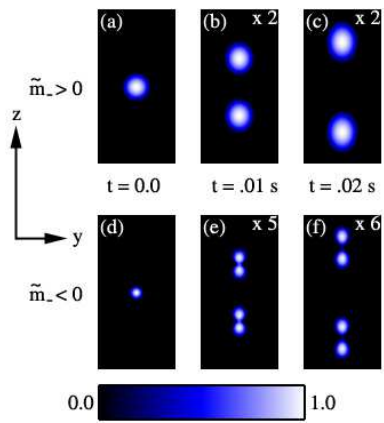} \caption{(Color online)
Splitting of atomic cloud for the positive effective mass case
(a-c) and negative mass case (d-f) after turning off the trap and
laser fields. The x $n$ label represents the value each panel has
been multiplied by to keep figure to scale.} \label{fig3}
\end{figure}

Figure \ref{fig3} displays numerical estimates of the splitting.
The time at which such splitting can be observed is found by
estimating the initial size of the atomic cloud, $R_F$. For system
with $N\approx10^{4\sim5}$ atoms and a trap frequency
$\omega\sim35$HZ, $R_F\approx230\mu m$ in the positive effective
mass case and $R_F\approx127.8\mu m$ when the effective mass is
negative, assuming $T<T_F$. As a result, for the positive
effective mass regime, one can verify that the two atomic clouds
will be fully separated after the system evolves by $\Delta
t\sim5ms$. For the negative mass regime, the time is about $\Delta
t\sim9ms$. It should be emphasized that measurement of the present
expansion dynamics needs to detect the density of atoms in the
hyperfine level $|b\rangle$ (or $|c\rangle$), rather than to
detect the pseudospin states $|\pm\rangle$ that are not
differentiable for the atomic system, thus the SO effect obtained
here is directly observable in experiments by direct imaging of
the separated atomic clouds.

Now we confirm the validity of the approximation
$M_0^2\gg\lambda_2^2\hbar^2(k^z_F)^2\gg\lambda_1^2\hbar^2(k^y_F)^2$.
The second inequality is always valid, since
$\lambda_2\gg\lambda_1$ in the large detuning case. Note that
$k_F^z$ can be calculated by
$k_F^z=\sqrt{\Upsilon+2m(\tilde{\varepsilon}_F+\hbar^2\Gamma\Upsilon^2/2m)/(\hbar^2\Gamma)}$
for $\tilde{m}_-<0$ and
$k_F^z=\sqrt{2\tilde{m}_-\tilde{\varepsilon}_F/\hbar^2}$ for
$\tilde{m}_->0$ . For the system composed of $N\sim10^4$ atoms and
a trap frequency $\omega=35$HZ, and using the previously employed
parameters, we find $k_F^z\approx1.41\times10^6m^{-1}$ with
$\lambda_2^2\hbar^2(k^z_F)^2/M_0^2\approx0.019\ll1$ for
$\tilde{m}_->0$, and $k_F^z\approx3.55\times10^6m^{-1}$ with
$\lambda_2^4\hbar^4(k^z_F)^4/M_0^4\approx0.028\ll1$ for
$\tilde{m}_-<0$ (note for the case $\tilde{m}_-<0$ the
approximation is up to the $k_z^4$ order). Thus, the first
inequality is also valid. Finally we estimate the TF approximation
that is considered in the calculation. TF approximation fails in a
small periphery region $R_F-\delta R<r<R_F$ of the atomic cloud
\cite{fermivelocity}. For the 2D fermi atom gas, one can find the
ratio of $\delta R$ to the atomic cloud size $R_F$ satisfies
$\delta R/R_F\sim N^{-1/2}$. Therefore such a small region can be
safely neglected for the case with a large number of atoms, say
$N>10^4$. Note this model can be readily extended to Bose-Einstein
condensate systems \cite{Galitski}, where, together with the
atom-atom interaction, the SO coupling may lead to intriguing new
physics.

In conclusion, we have proposed an experimental scheme to study SO
coupling effects for a cloud of a 2D trapped Fermi gas. Under
certain conditions, the optically induced SO coupling in atoms
leads to a spin-dependent effective mass which can be positive or
negative. In the expansion dynamics of the atomic cloud after
switching off the trap, it is shown that the initial atomic cloud
splits into two or four clouds moving oppositely depending on
tunable spin-orbit coupling parameters. The present scheme
provides an applicable way to directly observe the SO coupling in
cold Fermi atoms.

\begin{acknowledgments}
We gratefully acknowledge discussions with V. Galastki. This work
was supported by ONR under Grant No. ONR-N000140610122, by NSF
under Grant No. DMR-0547875, and by SWAN-NRI.  Jairo Sinova is a
Cottrell Scholar of the Research Corporation.
\end{acknowledgments}

\noindent

\begin{thebibliography}{99}

\bibitem{spintronics} S. A. Wolf {\it et al.}, Science {\bf 294}, 1488 (2001).

\bibitem{AHE} T. Jungwirth {\it et al.}, 
Phys. Rev. Lett. {\bf 88}, 207208 (2002).

\bibitem{SHE1} S. Murakami {\it et al.},
Science {\bf 301}, 1348 (2003); J. Sinova {\it et al.}, Phys. Rev.
Lett. {\bf 92}, 126603 (2004).

\bibitem{SHE2} Y. K. Kato {\it et al.}, Science, {\bf 306}, 1910 (2004);
J. Wunderlich {\it et al.}, Phys. Rev. Lett. {\bf 94}, 047204
(2005).

\bibitem{wilczek} F. Wilczek and A. Zee, Phys. Rev. Lett. 52, 2111 (1984).

\bibitem{atomgauge1} J. Ruseckas {\it et al.}, 
Phys. Rev. Lett. {\bf 95}, 010404 (2005).

\bibitem{atomgauge2} K. Osterloh {\it et al.}, 
Phys. Rev. Lett. {\bf 95}, 010403 (2005).

\bibitem{atomgauge3} X. -J. Liu {\it et al.},
Phys. Rev. Lett. {\bf 98}, 026602 (2007).

\bibitem{atomgauge4} S.-L. Zhu {\it et al.},
Phys. Rev. Lett. {\bf 97}, 240401 (2006).

\bibitem{atomgauge5} Yong Li, {\it et al.}, 
Phys. Rev. Lett. {\bf 99}, 130403 (2007).

\bibitem{spin-orbit} T. D. Stanescu, C. W. Zhang, and V. Galitski,
Phys. Rev. Lett. {\bf 99}, 110403 (2007).

\bibitem{Clark} J.Y. Vaishnav {\it et al.}, 
Phys. Rev. Lett. {\bf 100}, 153002 (2008).

\bibitem{niu1} A. M. Dudarev et al., Phys. Rev. Lett. {\bf 92}, 153005
(2004).

\bibitem{note} Note that in the large detuning case both the states
$|\chi_{D,B_1}\rangle$ decouple from the excited state $|a\rangle$
since $\sin\alpha\ll1$, and then the decay of them is negligible.

\bibitem{PHE} H. X. Tang {\it et al.},
Phys. Rev. Lett. {\bf 90}, 107201, (2003).

\bibitem{doublewell} Y. Shin, Experiments with Bose-Einstein Condensates in a Double-Well
Potential, Doctoral Thesis, MIT, 2005.

\bibitem{fermivelocity} D. A. Butts and D. S. Rokhsar,
Phys. Rev. A, {\bf 55}, 4346 (1997).


\bibitem{40K} B. DeMarco and D. S. Jin,
Science, {\bf 285}, 1703 (1999).

\bibitem{Granade} S. R. Granade {\it et al.}, 
Phys. Rev. Lett. {\bf 88}, 120405 (2002).

\bibitem{trap} R. Grimm {\it et al.}, Advances in Atomic, Molecular and Optical Physics Vol. 42, 95 (2000).

\bibitem{Galitski} T. Stanescu {\it et al.}, 
Phys. Rev. A {\bf 78}, 023616 (2008).

\end{thebibliography}
\end{document}